\documentclass[twocolumn,showpacs]{revtex4}
\usepackage[latin9]{inputenc}
\usepackage{amsmath}
\usepackage{amssymb}
\usepackage{esint}

\makeatletter
\@ifundefined{textcolor}{}
{%
 \definecolor{BLACK}{gray}{0}
 \definecolor{WHITE}{gray}{1}
 \definecolor{RED}{rgb}{1,0,0}
 \definecolor{GREEN}{rgb}{0,1,0}
 \definecolor{BLUE}{rgb}{0,0,1}
 \definecolor{CYAN}{cmyk}{1,0,0,0}
 \definecolor{MAGENTA}{cmyk}{0,1,0,0}
 \definecolor{YELLOW}{cmyk}{0,0,1,0}
}

\def\be{\begin{eqnarray} &&}

\def\ee{\end{eqnarray}}

\def\bew{\begin{widetext}}
\def\ew{\end{widetext}}

\makeatother

\begin{document}

\title{Comment on chiral symmetry restoration at finite density in
large-$N_{c}$ QCD}

\author{Hilmar Forkel}

\affiliation{Institut f\"ur Physik, Humboldt-Universit\"at zu Berlin, D-12489 Berlin,
Germany}
\begin{abstract}
In the article ``On chiral symmetry restoration at finite density
in large-$N_{c}$ QCD'' by Adhikari, Cohen, Ayyagari and Strother
{[}Phys. Rev. C \textbf{83}, 065201 (2011){]} the description of dense
nuclear matter by means of Skyrmions in hyperspherical unit cells
is severely criticized. We point out that this criticism is based
on invalid assumptions and therefore unwarranted.
\end{abstract}



\maketitle
Motivated by the suggested quarkyonic phase of dense matter at large
$N_{c}$ \cite{mcl07}, Adhikari et al. study in Ref. \cite{adh11}
conditions for chiral symmetry restoration at high baryon density
in Skyrme models \cite{zah86} and in large-$N_{c}$ QCD. Section
IV of their paper is devoted to the ``hypersphere approach\textquotedblright{}
\cite{man86,for89} which models certain aspects of dense matter (in
particular the transition to chiral restoration) by placing Skyrmions
in cells with the geometry of a three-dimensional sphere $S^{3}$.
Adhikari et al.'s discussion culminates in the claim that the results
of the $S^{3}$ approach are artifacts of a supposedly arbitrary choice
for the cell geometry from which not even qualitative physical insight
can be gained. The purpose of the following brief comment is to refute
this criticism by identifying a misconception and several incorrect
assumptions \cite{for12} which underly Ref. \cite{adh11}'s arguments.
(Hence no additional interpretation or justification for the $S^{3}$
approach will be required.)

The first of these assumptions is that the hypersphere approach was
intended to ``approximate a Skyrmion in the crystal''. The second,
more drastic one is that the cell shape is insignificant and motivated
by practical convenience only: \textquotedblleft{}Of course, this
geometry has no significance and was used for ease of computation\textquotedblright{}
(p. 10 of Ref. \cite{adh11}). In addition, Ref. \cite{adh11} assumes
that the $S^{3}$ approach is based on the premise ``that the principal
effect of putting a Skyrmion into a crystal is to restrict the space
over which it can spread'' and ``that\textbf{ }using a hypersphere
to restrict the volume of the Skyrmion acts generically like other
restrictions on its volume''.

Although the above assumptions were implied in Ref. \cite{adh11}
to be commonly accepted, they are (to the best of our knowledge) nowhere
stated in the hypersphere literature. In fact, especially the second
and third assumption are in plain contradiction with the unique symmetry
properties of the $S^{3}$ cell. These two assumptions deny precisely
the indispensable role of the $S^{3}$ geometry which, due to its
``chiral'' symmetry group SO$\left(4\right)\simeq$ SU$\left(2\right)\times$
SU$\left(2\right)$, made the approach promising in the first place
\cite{man86,for89,for12}. Indeed, already the pioneering papers \cite{man86,for89}
pointed out that \emph{only} in the $S^{3}$ geometry both the chirally-broken
and (in an averaged sense) restored phases can be modeled, that a
transition between them occurs at a critical energy and baryon density,
that the Skyrmion can attain its minimal energy only on $S^{3}$,
that parity doubling (including that of the former Goldstone pion
triplet) takes place in the restored phase as expected from complete
chiral restoration etc.. (The unique, curvature-generated interactions
in $S^{3}$ cells %
\footnote{Even in the quark-based Nambu-Jona-Lasinio model \cite{nam61}, when
put into $S^{3}$ cells, these curvature-induced interactions (instead
of just the reduced volume) generate a transition to a chirally restored
phase \cite{for92}. %
} were tentatively interpreted as modeling dense-matter-induced chiral
forces \cite{for89}.)

Furthermore, one should not regard hyperspherical cells as faithful
models of flat crystal unit-cells, several analogies and rather closely
shared results notwithstanding. In fact, the qualitative differences
between the two cell types were described in the literature (cf. Ref.
\cite{for12} for a summary) and strengthened the original view that
$S^{3}$ cells provide an independent description of dense matter
properties which may capture chiral features more completely than
the crystal approach, at least for $N_{c}<\infty$. (In particular,
one obviously cannot view the curved $S^{3}$ cells as regions of
flat space containing nuclear matter, or even as being in one-to-one
correspondence with unit cells of a crystal. Rather, the curvature-induced
interactions in $S^{3}$ were suggested to encode aspects of the dense
environment in a self-sufficient way, similar in spirit to ``analog-gravity''
models which proved useful in many areas of physics \cite{bar11}.)

By claiming that the cell geometry is insignificant and that all cell
shapes for Skyrmions should describe at least qualitatively similar
physics, Ref. \cite{adh11} therefore turns the logic of the hypersphere
approach on its head. This led to the erroneous conclusions that the
``evidence for chiral restoration ... was an artifact of the hyperspherical
geometry'' and even that ``the special properties of the geometry\textbf{
}...\textbf{ }make the (gained) intuition totally unreliable even
for qualitative issues associated with chiral symmetry breaking and
its possible restoration in the average sense''. In other words,
Ref. \cite{adh11} overlooks that the $S^{3}$ geometry was adopted
precisely for its unique symmetry properties which already the pioneering
papers recognized as indispensable for chiral restoration %
\footnote{This oversight becomes explicit in the calculation of Ref. \cite{adh11}'s
Sec. IV which is designed to show that a specific deformation of $S^{3}$
prevents (averaged) chiral restoration. Awareness of the fact that
any such deformation must break the crucial SO$\left(4\right)$ symmetry
of $S^{3}$ makes this calculation unnecessary since the result follows
directly from the symmetry arguments of Ref. \cite{man86}.%
}. Instead, Ref. \cite{adh11} claims the opposite, namely that ``this
geometry has no significance'' and ``that\textbf{ }using a hypersphere
to restrict the volume of the Skyrmion acts generically like other
restrictions on its volume'', and then suggests to discard the $S^{3}$
results for not complying with these unfounded claims.

(In fact, the claim that the principal effect of the cell shape, i.e.
of its topology and geometry, is just to restrict the cell volume
does not even hold for flat unit cells, including those of Skyrmion
crystals. The physics of the latter shows a remarkable sensitivity
to the cell shape encoded in the boundary conditions. Even the crucial
half-Skyrmion symmetry originates from a subtle change in the cell
form \cite{halfSkyrm} and not from a generic volume restriction.
The shape of unit cells in condensed-matter crystals often has a similarly
decisive physical impact.)

Finally, Ref. \cite{adh11} argues that all \emph{uniformly} spacially-averaged
chiral order parameters in large-$N_{c}$ QCD can simultaneously vanish
only if chiral symmetry is also restored in the conventional, local
sense, i.e. by a vanishing quark condensate. On the other hand, at
least superficially the former (uniformly averaged restoration) seems
to happen without the latter (local restoration) in $S^{3}$ cells.
There is no conflict, however, since spacial averaging over order
parameters on $S^{3}$ (on which in particular chiral restoration
in $S^{3}$ cells relies) differs qualitatively from the equal-weight
averaging in flat space %
\footnote{Analogous qualitative differences hold for the ``translational''
symmetry which the averaging procedure restores: ``translations''
on $S^{3}$ are elements of SO$\left(4\right)$ and do not commute,
for example, in contrast to their flat-space counterparts.%
} which underlies the arguments of Ref. \cite{adh11}. Indeed, there
is no reason for spacial averaging over an $S^{3}$ cell to translate
into uniform spacial averaging over some flat-space region or configuration.
Hence, Ref. \cite{adh11}'s arguments imply no contradiction between
the $S^{3}$-averaged hypersphere results and uniformly averaged large-$N_{c}$
QCD results. 

To summarize: Ref. \cite{adh11}'s criticism of the hypersphere approach
to dense matter is based on invalid assumptions and therefore unwarranted.
The $S^{3}$ cell geometry, in particular, is uniquely dictated by
chiral symmetry and thus an indispensable part of the approach. The
qualitative differences between spacial averaging in flat space and
in curved cells, furthermore, prevent any conflict with general arguments
of Ref. \cite{adh11} regarding ``uniformly flat-space averaged''
chiral restoration in large-$N_{c}$ QCD. 
\begin{acknowledgments}
We acknowledge extensive correspondence and discussions with Thomas
Cohen.\end{acknowledgments}

\end{document}